\documentstyle[aps,floats,psfig]{revtex}

\begin{document}
\psfigurepath{.:plot:figure}
\twocolumn[\hsize\textwidth\columnwidth\hsize\csname @twocolumnfalse\endcsname
\bibliographystyle{unsrt}
\preprint{Ce218b.tex, \today}
\title{
Magnetic structure of heavy fermion Ce$_2$RhIn$_8$
}
\author{Wei Bao, 
P.G. Pagliuso, J.L. Sarrao, 
J.D. Thompson and Z. Fisk$^{*}$}
\address{Los Alamos National Laboratory, Los Alamos, NM 87545}
\author{J. W. Lynn}
\address{NIST Center for Neutron Research, National Institute of Standards 
and Technology, Gaithersburg, MD 20899}
\date{\today}
\maketitle
\begin{abstract}
The magnetic structure of the heavy fermion antiferromagnet
Ce$_2$RhIn$_8$ is determined using neutron diffraction. 
It is a collinear antiferromagnet with a staggered moment of 
0.55(6)$\mu_B$ per Ce at 1.6~K, tilted 38$^o$ from the tetragonal
$c$ axis. In spite of its layered crystal
structure, the phases for the magnetic
moments are the same as those in the cubic parent 
antiferromagnet CeIn$_3$. 
This suggests that the cubic CeIn$_3$ building blocks have a stronger
influence on magnetic correlations than intervening layers, which
gives the material its apparent two-dimensional lattice structure
and renders CeRhIn$_5$ an incommensurate antiferromagnet.
\end{abstract}
\vskip2pc]
%\pacs{PACS numbers: }

\narrowtext

Superconducting heavy fermion materials belong to a special class
of correlated electron systems where unconventional superconductivity
may be mediated by magnetic fluctuations\cite{ott_fisk}.
Until recently, there were only five U-based heavy
fermion materials showing superconductivity at ambient pressure\cite{bobh}
in addition to the original heavy fermion 
superconductor CeCu$_2$Si$_2$\cite{steg}.
Three Ce-based heavy fermion materials isostructural to CeCu$_2$Si$_2$
and cubic CeIn$_3$ become superconductors under pressure.
Recently, superconductivity has been discovered in a new structure class of
heavy fermion materials with chemical formulas Ce$M$In$_5$.
While the $M=$Rh member superconducts below 2.1~K under 17kbar\cite{hegger},
the $M=$Ir and Co members superconduct below 0.4~K and 2.3~K, respectively,
at ambient pressure\cite{joeIr,joeCo}. 
The high superconducting transition temperatures
of the new materials hold the record for heavy fermion superconductors.
Thermodynamic and transport measurements at low temperature 
are consistent with unconventional superconductivity in which there
are lines of nodes in the superconducting gap\cite{roman}.

Because CeIn$_3$ and Ce$M$In$_5$ belong to the
Ce$_nM_m$In$_{3n+2m}$ family of structures, they present a unique
opportunity for investigating the influence of systematic
structure modifications on the superconducting and magnetic 
properties\cite{joe}. In particular,
it is interesting to compare CeIn$_3$ and CeRhIn$_5$, which
are the $n=\infty$ and $n=1$ members of the Ce$_n$RhIn$_{3n+2}$
sub-family, and can be viewed as periodic stacking of
$n$-layers of CeIn$_3$ on a layer of $M$In$_2$\cite{russ,nstru}.
Both are antiferromagnetic at ambient pressure with 
$T_N= 10$~K for CeIn$_3$\cite{ceinm} and $T_N=3.8$~K
for CeRhIn$_5$\cite{hegger,nqr}. Both become superconductors 
when subjected to pressure\cite{hegger,cmbr}, with
the superconducting transition temperature of CeRhIn$_5$ being one 
order of magnitude higher that that for CeIn$_3$.
This raises a fundamental question about the role of the
intervening $M$In$_2$ layers on both the superconductivity
and antiferromagnetism. Our study of Ce$_2$RhIn$_8$, which is the
$n=2$ member of this heavy fermion sub-family, is intended to
shed light on this question by changing the ratio of the CeIn$_3$ and
RhIn$_2$ layers.

\begin{figure}[bt]
\centerline{
\psfig{file=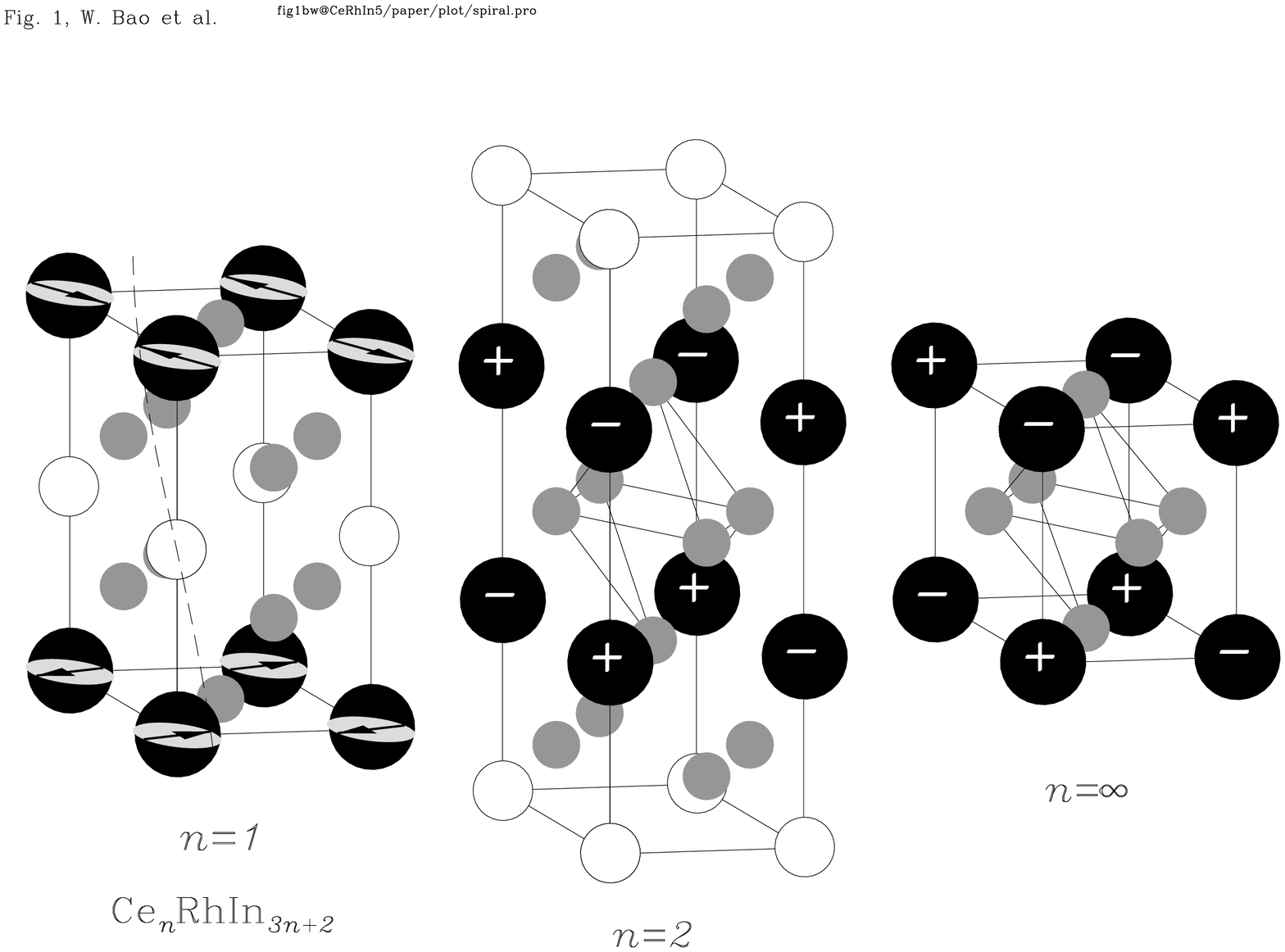,width=\columnwidth,angle=90,clip=}}
\caption{The magnetic structure of Ce$_2$RhIn$_8$ ($n=2$) 
in a structural unit cell
is shown together with CeRhIn$_5$ ($n=1$)\protect\cite{bao00a}
and CeIn$_3$ ($n=\infty$)\protect\cite{ssc,cein}. The magnetic moment 
is 0.55$\mu_B$ per Ce in Ce$_2$RhIn$_8$ and it points
38$^o$ from the $c$ axis.
The solid circle denotes Ce, the shaded circle
In, and the open circle Rh.
The disk for CeRhIn$_5$ denotes the plane in which the
ordered moment rotates.
}
\label{mstr}
\end{figure}

Antiferromagnetic structures for both CeIn$_3$ and CeRhIn$_5$ have
been determined previously. Cubic CeIn$_3$ has a simple commensurate
magnetic order with wave vector (1/2,1/2,1/2)
below its N\'{e}el temperature. The staggered 
magnetic moment is 0.48-0.65$\mu_B$ per Ce\cite{ssc,cein}. 
In contrast, magnetic moments of Ce ions in tetragonal 
CeRhIn$_5$ form an incommensurate transverse
spiral below $T_N=3.8$~K, with wave vector 
(1/2,1/2,0.297)\cite{bao00a}(refer to Fig.~\ref{mstr}).
The staggered moment of 0.37 $\mu_B$ per Ce at 1.4~K is smaller
than that for CeIn$_3$. In this paper, we report the magnetic structure
for Ce$_2$RhIn$_8$, which orders at $T_N=2.8$~K\cite{joe}. 
The commensurate antiferromagnetic structure 
for this $n=2$ material closely resembles that for 
CeIn$_3$ ($n=\infty$), rather than the magnetic spiral
of CeRnIn$_5$ ($n=1$). This suggests a strong influence of the cubic
CeIn$_3$ structural unit on magnetic correlations in this family of
heavy fermion materials.

Single crystals of Ce$_2$RhIn$_8$ were grown from an In flux.
They crystallize in the tetragonal Ho$_2$CoGa$_8$ structure 
(space group \#123, P4/mmm)\cite{russ}, 
with lattice parameters $a=4.665\AA$ and $c=12.244\AA$ 
at room temperature.
The sample used in this study was a well-faceted rectangular plate
of dimension $\sim4\times 4\times 0.7$~mm and weight of 88 mg. 
The largest surface is the (001) plane.
Neutron diffraction experiments were performed at NIST 
using the thermal triple-axis spectrometer BT2 in a two-axis mode.
The horizontal collimations were 60-40-40-open. 
Neutrons with incident energy $E=35$ meV were 
selected using the (002) reflection
of a pyrolytic graphite (PG) monochromator. The neutron penetration
length at this energy is 1.8 mm, which is substantially longer
than the thickness of the sample. No rocking-angle dependent
absorption was noticed. PG filters of total 9
cm thickness were used to remove higher order neutrons. 
The sample temperature was regulated by a top
loading pumped He cryostat.

\begin{figure}[bt]
\centerline{
\psfig{file=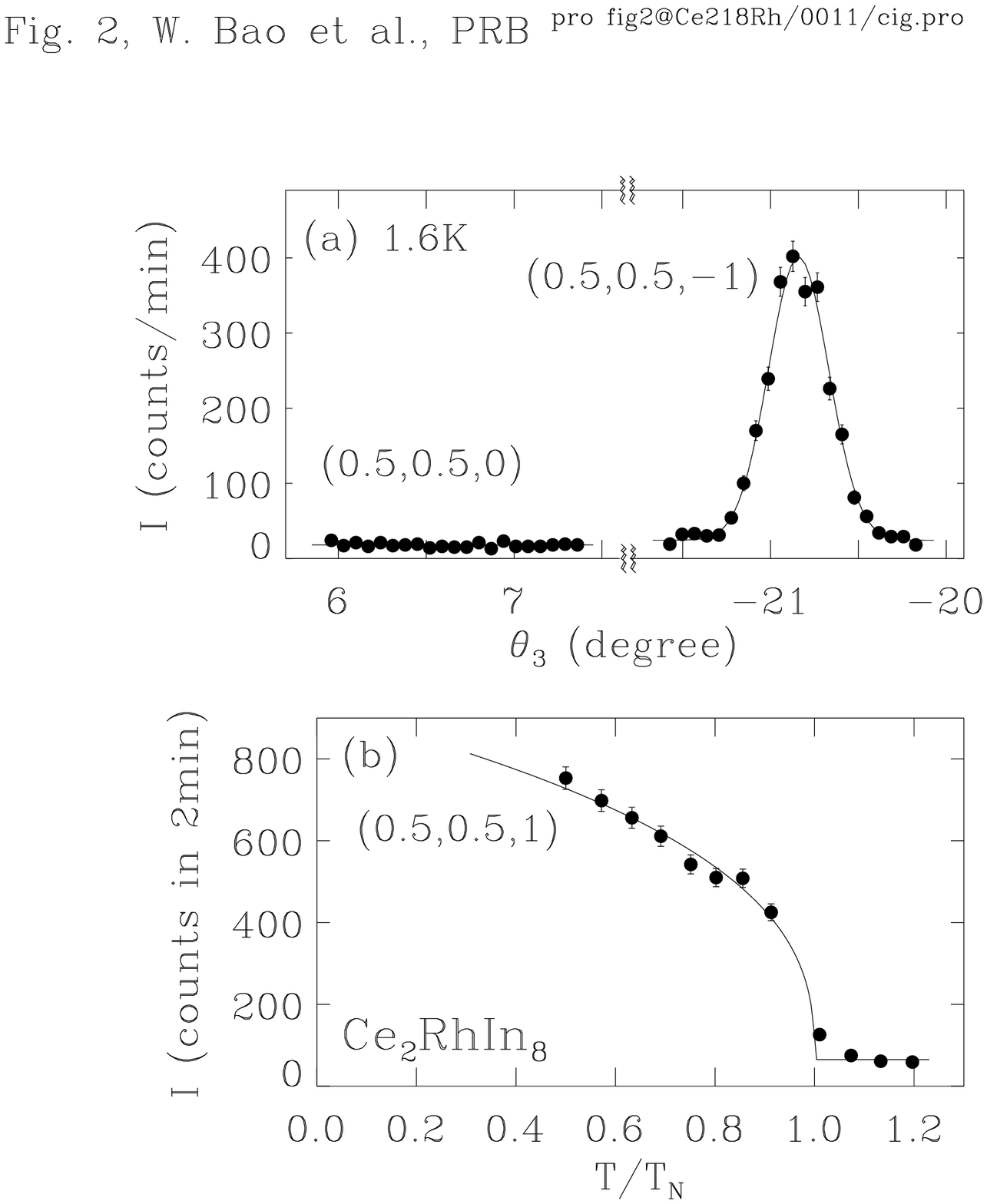,width=\columnwidth,angle=0,clip=}}
\caption{(a) Elastic rocking scans through magnetic Bragg points 
(1/2,1/2,0) and (1/2,1/2,$-1$) at 1.6~K.
The (1/2,1/2,0) is forbidden for this magnetic structure.
(b) Intensity of (1/2,1/2,1) as a function of T/T$_N$.
The N\'{e}el temperature is 2.8~K.
}
\label{scan}
\end{figure}

Temperature-dependent magnetic Bragg peaks were found at
($m$/2,$n$/2,$l$) with $m$ and $n$ odd integers and
$l$ {\it non-zero} integers. This corresponds to a magnetic unit cell
that doubles the structural unit cell in the basal plane
and contains four magnetic Ce ions.
Rocking scans at (1/2,1/2,0) and (1/2,1/2,$\overline{1}$), 
taken at 1.6~K, are shown in 
Fig.~\ref{scan}(a). The intensity of the (1/2,1/2,1) peak
is shown in Fig.~\ref{scan}(b) as the square of the order parameter
of the magnetic phase transition.
Integrated intensities of magnetic Bragg peaks
from such rocking scans are normalized to structural Bragg peaks (001), (002),
(003), (005), (006) and (220) to yield magnetic
scattering cross sections, $\sigma({\bf q})=I({\bf q}) \sin(\theta_4)$,
in absolute units (see Table~\ref{mlist}). 
\begin{table}
\caption{Magnetic Bragg intensity, $\sigma_{obs}$,
defined in Eq.~(\ref{eq_cs}), 
observed at
1.6~K in units of $10^{-3}$ barns per Ce$_2$RhIn$_8$.
The theoretical intensity, $\sigma_{cal}$, is calculated using Eq.~(\ref{eq_now}) and (\ref{eq_pol}) with $\beta=52^o$
and $M=0.55\mu_B$/Ce.
}
\label{mlist}
\begin{tabular}{cdd}
${\bf q}$ & $\sigma_{obs}$ & $\sigma_{cal}$\\
\hline
(     0.5     0.5     $-1$  ) &   52(1)  & 46.2\\
(     0.5     0.5      0  ) &   0.0(3) & 0.0\\
(     0.5     0.5      1  ) &   49(1)  & 46.2\\
(     0.5     0.5      2  ) &   19(1)  & 18.9\\
(     0.5     0.5      3  ) &   6.4(4) &  5.5\\
(     0.5     0.5      4  ) &   21(1)  & 23.2\\
(     0.5     0.5      6  ) &   7.5(7) & 7.6\\
(     1.5     1.5      0  ) &   0.0(8) & 0.0\\
(     1.5     1.5      1 )  &   18(1)  & 24.8 \\
\end{tabular}
\end{table}
In such units, the magnetic cross section is\cite{neut_squire}
\begin{equation}
\sigma({\bf q})=\left(\frac{\gamma r_0}{2}\right)^2
	\langle M\rangle^2 \left|f(q)\right|^2 
	\sum_{\mu,\nu}(\delta_{\mu\nu}
	-\widehat{\rm q}_{\mu}\widehat{\rm q}_{\nu})
	{\cal F}^*_{\mu}({\bf q}){\cal F}_{\nu}({\bf q}),
\label{eq_cs}
\end{equation}
where $(\gamma r_0/2)^2=0.07265$~barns/$\mu_B^2$, $M$ is 
the staggered moment of the Ce ion, $f(q)$ the Ce$^{3+}$ magnetic
form factor\cite{formf_ce}, $\widehat{\bf q}$ the unit vector of ${\bf q}$,
and ${\cal F}_{\mu}({\bf q})$ the $\mu$th
Cartesian component of magnetic structure factor per Ce$_2$RhIn$_8$.

Forbidden peaks at ($m$/2,$m$/2,0) provide an important clue
to the magnetic structure of Ce$_2$RhIn$_8$. They could be due to magnetic 
moments aligning along the [110] direction. However, magnetic
twinning in this tetragonal material will yield finite intensities
at these reciprocal points. Another, more reasonable, cause
is that the nearest-neighbor magnetic moments along the $c$ axis are antiparallel. The phase between the next nearest-neighbor
magnetic moments along the $c$ axis and
the phases of magnetic moments in a basal layer are already determined
by the magnetic wave vector.
This yields a collinear antiferromagnetic structure
(refer to Fig.~\ref{mstr}) with magnetic 
cross sections per Ce$_2$RhIn$_8$
\begin{equation}
\sigma({\bf q})=4\left(\frac{\gamma r_0}{2}\right)^2
	\langle M\rangle^2 \left|f(q)\right|^2 
	\langle 1-(\widehat{\bf q}\cdot \widehat{\bf s})^2\rangle
	\sin^2(l\epsilon),
\label{eq_now}
\end{equation}
where $2\epsilon=0.38 c$ is the separation between the nearest-neighbor
Ce ions along the $c$ axis, $\widehat{\bf s}$ is the unit vector
of the magnetic moment, and the average, 
$\langle 1-(\widehat{\bf q}\cdot \widehat{\bf s})^2\rangle$,
is over magnetic domains.

\begin{figure}[bt]
\centerline{
\psfig{file=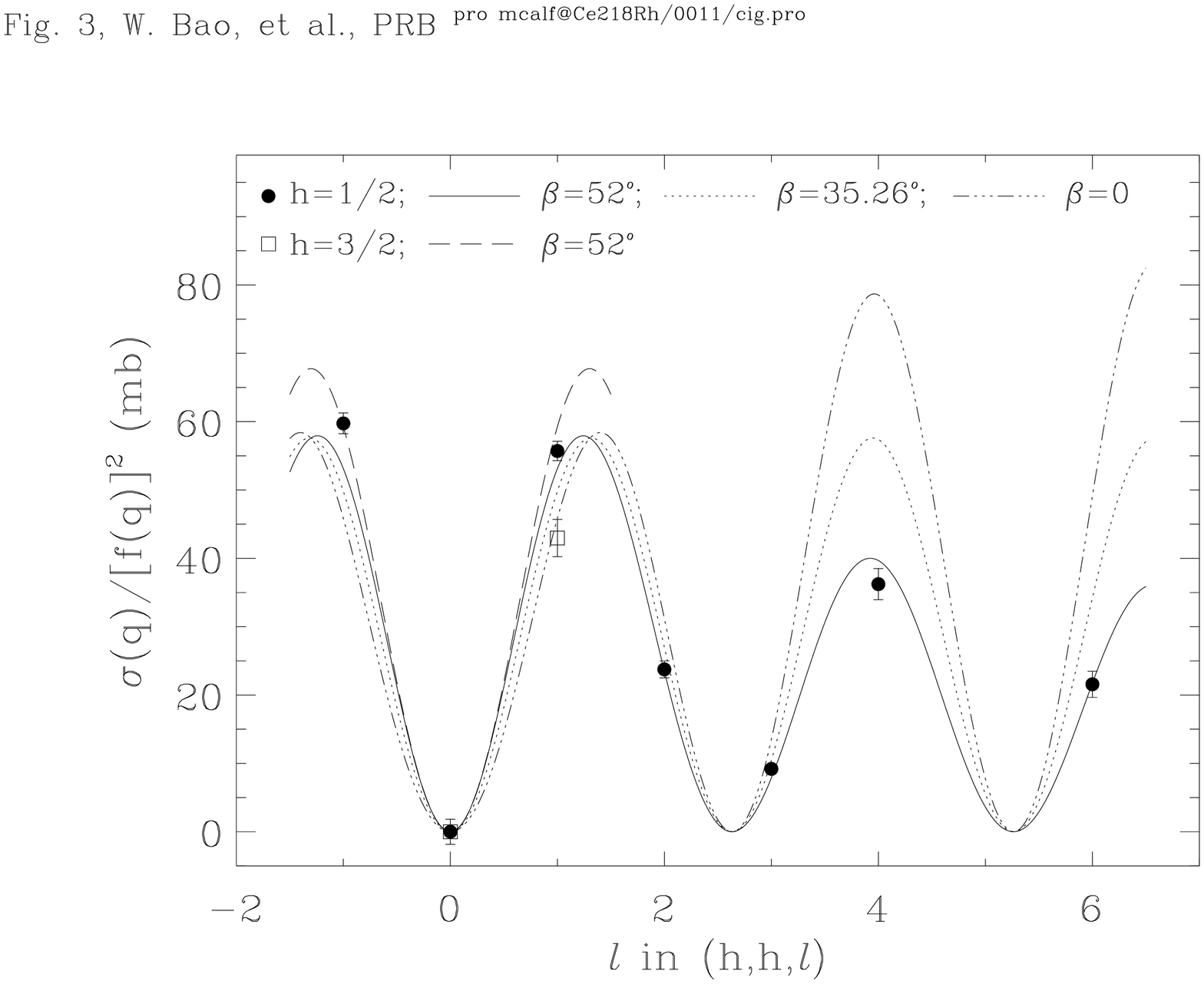,width=\columnwidth,angle=90,clip=}}
\caption{The $l$ dependence of the magnetic cross-section, $\sigma$, divided by 
the form factor $\left|f(q) \right|^2$. 
The theoretical curves are calculated using
Eq.~(\ref{eq_now}) and (\ref{eq_pol}) with $M=0.55\mu_B$ and
various values specified in the figure for the moment tilt angle, $\beta$.} 
\label{polf}
\end{figure}

Fig.~\ref{polf} shows 
$\sigma_{obs}({\bf q})/\left|f(q) \right|^2\sim
\langle 1-(\widehat{\bf q}\cdot \widehat{\bf s})^2\rangle\sin^2(l\epsilon)$ 
as a function
of the $l$ index of {\bf q}. The structure factor, $4\sin^2(l\epsilon)$,
not only accounts for the forbidden $l=0$ magnetic peaks, but it also
accounts for the strong oscillation of $\sigma_{obs}$ as a function of $l$. 
The remaining, smooth $l$ dependence is
to be accounted for by the polarization factor 
$\langle 1-(\widehat{\bf q}\cdot \widehat{\bf s})^2\rangle$.

Denote the angle between {\bf q} and the basal
plane as $\alpha$, and the angle between the basal plane
and the magnetic moment as $\beta$. Assuming equal
occupations among magnetic twins, we have
\begin{equation}
\langle 1-(\widehat{\bf q}\cdot \widehat{\bf s})^2\rangle
= 
1-\frac{\cos^2\alpha \cos^2\beta+2\sin^2\alpha\sin^2\beta}{2}.
\label{eq_pol}
\end{equation}
For a magnetic moment lying in
the basal plane, $\beta=0$, which is the case for CeRhIn$_5$\cite{bao00a,nqr},
the polarization factor varies too much. The resulting theoretical
curve does not fit the data (refer to the dot-dashed line in Fig.~\ref{polf}). 
For $\beta=35.26^o$, which corresponds to {\bf s}
in the $\langle 111\rangle$ directions in a cubic system,
the polarization factor averages to a constant, 2/3. The resulting theoretical
curve is a better fit (refer to the dotted line 
in Fig.~\ref{polf}) than that for $\beta=0$,
but it is still not satisfactory. 
The best least-squares fit (refer to the solid and 
dashed lines for $h=k=1/2$ and
$h=k=3/2$ respectively) yields $\beta=52(2)^o$.
The staggered magnetic moment is determined at 1.6~K
to be $M=0.55(6) \mu_B$ per Ce.

Having determined the magnetic structure of Ce$_2$RhIn$_8$,
now we consider the systematics relating the magnetic structure and
lattice structure in Ce$_n$RhIn$_{3n+2}$ (see Fig.~\ref{mstr}).
In the $a$-$b$ plane, the magnetic moments of the Ce ions 
form a square lattice,
surrounded by In ions, in all three materials. They all are 
simple, nearest-neighbor antiferromagnets in the basal plane. In 
CeRhIn$_5$, this Ce antiferromagnetic plane alternates with
the RhIn$_2$ layer. Magnetic correlations across the RhIn$_2$ layer are
incommensurate, with neighboring magnetic moments
being rotated by 107$^o$\cite{bao00a}.
The local structure environment in the vertical $a$-$c$ or the $b$-$c$
plane within the CeIn$_3$ double layer in Ce$_2$RhIn$_8$ is very similar to
that in the basal layer. The same nearest-neighbor antiferromagnetic
correlations exist in the double layers. It is interesting that now
across the RhIn$_2$ layer the Ce moments are antiparallel 
instead of rotated by 107$^o$. The insertion of the RhIn$_2$
layers between CeIn$_3$ bilayers, thus, does not modify 
the magnetic order relative to cubic CeIn$_3$. 
This suggests CeRhIn$_5$ as a unique member of
the Ce$_n$RhIn$_{3n+2}$ family, and the $n\ge 2$ members are likely
to be magnetically similar to cubic CeIn$_3$.
Searching for heavy fermion materials with two-dimensional magnetism
seems more profitable if one could find a Ce$M_m$In$_{3+2m}$ 
structure family,
where $m$ $M$In$_2$ layers separate a single CeIn$_3$ layer.

Another interesting difference between the $n=1$ and the $n=2$
materials concerns the magnetic moment orientation. In CeRhIn$_5$,
the moments rotate in the $a$-$b$ plane, indicating a $XY$ type magnetic
anisotropy. In Ce$_2$RhIn$_8$, the
magnetic moments point 52$^o$ from the basal plane. 
Different local anisotropic fields, together with isotropic
exchange and crystal fields, likely contribute to the different
magnetic structures in the two materials\cite{jjarm}.
We also notice that the staggered moment of Ce$_2$RhIn$_8$ is
comparable to that of CeIn$_3$, while $T_N$ of Ce$_2$RhIn$_8$
is closer to that of CeRhIn$_5$. The $T_N/\langle M\rangle^2$ is 
the smallest for Ce$_2$RhIn$_8$ in the group. 
Antiferromagnetic transition has been studied in 
isostructural Nd$_nM$In$_{3n+2}$ ($M$=Rh,Ir and $n=1$,2)
and crystal field effects have been emphasized\cite{pgNd}.
A detailed understanding of the magnetic interaction in 
these materials is essential
in pursuing magnetic origin of unconventional
superconductivity in these heavy fermion
metals.

In conclusion, we find the magnetic structure of Ce$_2$RhIn$_8$
to be closely related to that of cubic CeIn$_3$. The staggered
moment is 0.55(6) $\mu_B$ per Ce at 1.6~K and it points 52$^o$
from the $a$-$b$ plane. Understanding the different magnetic structures
in Ce$_2$RhIn$_8$ and CeRhIn$_5$ may help us understand the enormous
enhancement in the superconducting transition temperature of
CeRhIn$_5$ over its cubic relative.

We thank M. E. Zhitomirsky and M. F. Hundley for useful discussions.
WB thanks D. C. Dender for technical assistance at NIST.
Work at Los Alamos was performed under the auspices of the US Department
of Energy. ZF gratefully acknowledges NSF support at FSU. PGP acknowledges
FAPESP for partial support.

\end{document}